\documentclass[twocolumn,showpacs,showkeywords]{revtex4-1}%
\usepackage{amsmath}
\usepackage{amssymb}
\usepackage{amsfonts}
\usepackage{amssymb}
\usepackage{graphicx}
\usepackage{txfonts}
\usepackage[linkcolor=blue,anchorcolor=black,citecolor=blue]{hyperref}%
\providecommand{\U}[1]{\protect\rule{.1in}{.1in}}

\begin{document}
\title{Improving spin-based noise sensing by adaptive measurements}
\author{Yi-Hao Zhang}
\author{Wen Yang}
\email{wenyang@csrc.ac.cn}
\affiliation{Beijing Computational Science Research Center, Beijing 100193, China}

\begin{abstract}
Localized spins in the solid state are attracting widespread attention as
highly sensitive quantum sensors with nanoscale spatial resolution and
fascinating applications. Recently, adaptive measurements were used to improve
the dynamic range for spin-based sensing of deterministic Hamiltonian
parameters. Here we explore a very different direction -- spin-based adaptive
sensing of random noises. First, we identify distinguishing features for the
sensing of magnetic noises compared with the estimation of deterministic
magnetic fields, such as the different dependences on the spin decoherence,
the different optimal measurement schemes, the absence of the modulo-2$\pi$
phase ambiguity, and the crucial role of adaptive measurement. Second, we
perform numerical simulations that demonstrate significant speed up of the
characterization of the spin decoherence time via adaptive measurements. This
paves the way towards adaptive noise sensing and coherence protection.

\end{abstract}
\maketitle

\section{Introduction}

Localized electronic spins in the solid state, such as nitrogen-vacancy
centers in diamond \cite{RondinRPP2014}, phosphorus donors
\cite{PlaNature2012}, silicon vacancy in SiC \cite{WidmannNatMater2014}, and
single rare-earth ion in yttrium aluminium garnet \cite{KolesovNatCommun2012},
are attracting widespread attention as highly sensitive quantum sensors
\cite{DegenRMP2017} with nanoscale spatial resolution
\cite{BalasubramanianNature2008,BalasubramanianNatMater2009,MaletinskyNatNano2012,GrinoldsNatPhys2013,GrinoldsNatNano2014,MullerNatComm2014}
and fascinating applications in condensed matter physics, materials science,
and biology. The coherent Larmor precession of the spin reveals deterministic
magnetic signals
{{\cite{ChernobrodJAP2005,DegenAPL2008,TaylorNatPhys2008,MazeNature2008,WangNJP2017}%
, while the decoherence of the spin reveals }}random{{ magnetic noises }%
}\cite{SchoelkopfBook2003,SousTAP2009,HallPRL2009,HallPNAS2010} and other
quantum objects
\cite{ZhaoNatNano2011,ZhaoNatNano2012,KolkowitzPRL2012,TaminiauPRL2012,LondonPRL2013,StaudacherScience2013,MaminScience2013,ShiNatPhys2014}%
. By tracking the noises back to the environment, the localized spin can
further reveal the structure and many-body physics of the environments, such
as quantum criticality \cite{QuanPRL2006,ChenNJP2013} and partition functions
in the complex plane
\cite{WeiPRL2012,WeiSciRep2014,PengPRL2015,WeiSciRep2015,HeylPRL2013} and the
quantum work spectrum \cite{DornerPRL2013,MazzolaPRL2013,BatalhaoPRL2014}. In
these developments, the key challenge is to improve the sensing precision. For
this purpose, dynamical decoupling techniques -- originally developed for
protecting qubits from decoherence -- have been adapted for sensing
alternating signals \cite{KotlerNature2011,LangePRL2011}, noises
\cite{LangeScience2010,AlvarezPRL2011,MedfordPRL2012,BarGillNatCommun2012,MuhonenNatNano2014}%
, and other quantum objects
\cite{ZhaoNatNano2011,ZhaoNatNano2012,KolkowitzPRL2012,TaminiauPRL2012,LondonPRL2013,StaudacherScience2013,MaminScience2013,ShiNatPhys2014,ZhaoPRA2014a,ZhaoPRA2014,MaPRA2015,CasanovaPRA2015,WangPRB2016b,XiaoNJP2016,WangNC2017}%
. Other techniques include rotating-frame magnetometry
\cite{CaiNJP2013,YanNatComm2013,LoretzPRL2013}, Floquet spectroscopy
\cite{LangPRX2015}, two-dimensional spectroscopy \cite{BossPRL2016,MaPRA2016},
correlative measurements \cite{LaraouiNatCommun2013}, axillary quantum memory
\cite{ZaiserNC2016}, and compressive sensing
\cite{ShabaniPRL2011,BossScience2017}.

Recently, there were growing interest in using adaptive measurements to
mitigate the modulo-2$\pi$ phase ambiguity and hence improve the dynamic range
of spin-based quantum sensing
\cite{SergeevichPRA2011,SaidPRB2011,WaldherrNatNano2012,NusranNatNano2012,BonatoNatNano2016,StenbergPRL2014}%
. However, previous works focus on \textit{deterministic }Hamiltonian\textit{
}parameters that drive the \textit{unitary} evolution of the spin quantum
sensor, leaving a large, important family of tasks unexplored -- the
spin-based quantum sensing of \textit{random} noises that drive the
\textit{non-unitary} decoherence of the spin. It is important to identify the
distinctions of spin-based noise sensing compared with the spin-based
Hamiltonian parameter estimation, and further provide feasible methods to
improve the key figure of merit -- the sensing precision.

In this work, we explore theoretically the role of adaptive measurement in
spin-based sensing of magnetic noises. First, our general analysis identifies
a series of distinguishing features for sensing a \textit{random} magnetic
field (i.e., magnetic noises) compared with the estimation of a
\textit{deterministic} magnetic field (which is a paradigmatic Hamiltonian
parameter), including the different dependences on the spin decoherence, the
different optimal measurement schemes, and the absence of the modulo-2$\pi$
phase ambiguity. Moreover, optimizing noise sensing requires knowledge about
the unknown noises to be estimated, so adaptive measurements are crucial for
improving the sensing precision. By contrast, in the estimation of
deterministic magnetic fields, adaptive measurements are usually alternatives
to non-adaptive schemes for mitigating the modulo-2$\pi$ phase ambiguity and
hence improving the dynamic range\textit{ }%
\cite{SaidPRB2011,WaldherrNatNano2012} and non-adaptive measurements can even
outperform adaptive ones in some cases \cite{SaidPRB2011}. Second, we perform
numerical simulations and demonstrate that using adaptive measurements can
speed up significantly the estimation of the spin decoherence time. These
results pave the way towards spin-based adaptive sensing of noises. Since
rapid characterization of decoherence allows us to design efficient schemes to
suppress the decoherence, these results are also relevant to quantum computation.

This rest of this paper is organized as follows. In Sec. II, we outline the
basic steps of a general adaptive measurement, leaving a detailed introduction
to every step in Appendices A-E. In Sec. III, we analyze general spin-based
sensing of magnetic noises and identify its distinguishing features. In Sec.
IV, we perform numerical simulations for the adaptive estimation of the spin
decoherence time. In Sec. IV, we draw the conclusion.

\section{Adaptive quantum parameter estimation}

\begin{figure}[ptb]
\includegraphics[width=\columnwidth,clip]{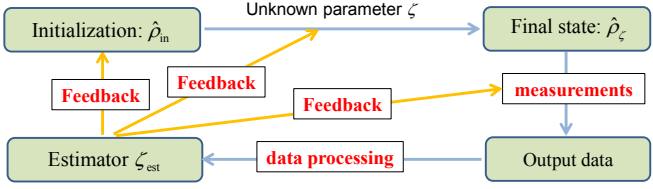}\caption{General
framework of adaptive quantum parameter estimation.}%
\label{G_FRAMEWORK}%
\end{figure}

A general parameter estimation protocol using a quantum system to estimate an
unknown, real parameter $\zeta$ consists of three steps (Fig.
\ref{G_FRAMEWORK}):

\leftmargini=3mm

\begin{enumerate}
\item The quantum system is prepared into certain (usually nonclassical)
initial state $\hat{\rho}_{\mathrm{in}}$ and then undergoes certain $\zeta
$-dependent evolution into a final state $\hat{\rho}_{\zeta}$. This step
encodes the information about $\zeta$ into the final state $\hat{\rho}_{\zeta
}$ of the quantum system. The information contained in $\hat{\rho}_{\zeta}$ is
quantified by the quantum Fisher information (QFI)$\ \mathcal{F}(\zeta)$.

\item The quantum system undergoes a measurement, which produces an outcome
according to certain probability distribution. In this step, the quantum
Fisher information $\mathcal{F}(\zeta)$ contained in $\hat{\rho}_{\zeta}$ is
transferred into the classical information in the measurement outcome. The
information contained in each outcome is quantified by the classical Fisher
information (CFI) $F(\zeta)$, which obeys
\begin{equation}
F(\zeta)\leq\mathcal{F}(\zeta). \label{FCFQ}%
\end{equation}

\item Steps 1-2 are repeated $N$ times and the $N$ outcomes are processed to
yield an estimator $\zeta_{\mathrm{est}}$ to the unknown parameter $\zeta$. In
this step, the total CFI $NF(\zeta)$ contained in the $N$ outcomes is
converted to the estimation precision, as quantified by the statistical error
of the estimator:
\begin{equation}
\delta\zeta\equiv\sqrt{\langle(\zeta_{\mathrm{est}}-\zeta)^{2}\rangle
},\label{UNCERTAINTY}%
\end{equation}
where $\langle\cdots\rangle$ denotes the average over a lot of estimators
obtained by repeating steps 1-3 many times. For unbiased estimators obeying
$\langle\zeta_{\mathrm{est}}\rangle=\zeta$, the precision $\delta\zeta$ is
fundamentally limited by the inequality%
\begin{equation}
\delta\zeta\geq\frac{1}{\sqrt{NF(\zeta)}}\Leftrightarrow(\delta\zeta)^{-2}\leq
NF(\zeta),\label{CRB}%
\end{equation}
known as the Cram\'{e}r-Rao bound \cite{HelstromBook1976,BraunsteinPRL1994}.
\end{enumerate}

For optimal performance, it is necessary to optimize each step of the above
initialization-evolution-measurement cycle. In step 1, the initial state
$\hat{\rho}_{\mathrm{in}}$ and the evolution process should be optimized to
maximize $\mathcal{F}(\zeta)$. In step 2, appropriate measurements should be
designed to convert all the QFI contained in $\hat{\rho}_{\zeta}$ into the CFI
contained in the measurement outcome, so that $F(\zeta)$ attains its maximum
value $\mathcal{F}(\zeta)$ allowed by Eq. (\ref{FCFQ}). In step 3, optimal
unbiased estimators should be used to convert all the CFI $NF(\zeta)$
contained in the $N$ outcomes into the useful information $(\delta\zeta)^{-2}$
contained in the estimator. For example, for large $N$, the Bayesian estimator
or the maximum likelihood estimator are unbiased and can saturate the
Cram\'{e}r-Rao bound Eq. (\ref{CRB}).

In step 1 and step 2, $\mathcal{F}(\zeta)$ and $F(\zeta)$ may depends on
$\zeta$, so the optimization for maximal $\mathcal{F}(\zeta)$ and $F(\zeta)$
requires knowledge about the true value (denoted by $\zeta_{\mathrm{true}}$)
of the unknown parameter $\zeta$. A possible solution is adaptive measurements
\cite{NielsenJPAMG2000,FujiwaraJPA2006}, i.e., using the measurement outcomes
of previous initialization-evolution-measurement cycles to refine our
knowledge about $\zeta$ and then use this knowledge to optimize the next
cycle. In step 2-3, the probability distribution of the measurement outcome as
a function of the unknown parameter $\zeta$ may be periodic, making it
impossible to identify a unique estimator. This ambiguity problem is commonly
encountered in estimating deterministic Hamiltonian parameters and can be
mitigated by using either non-adaptive or adaptive measurements
\cite{SaidPRB2011}. In this work, we explore the spin-based sensing of random
noises and show that the ambiguity problem is absent, while the dependence of
$\mathcal{F}(\zeta)$ and $F(\zeta)$ on $\zeta_{\mathrm{true}}$ makes adaptive
measurements critical for improving the sensing precision.

Our subsequent discussions are based on the general adaptive measurement
protocol in Fig. \ref{G_FRAMEWORK}, which involve many important concepts and
techniques, such as the QFI, the CFI, optimal unbiased estimators (such as the
Bayesian estimator and the maximum likelihood estimator), the Cram\'{e}r-Rao
bound, and adaptive measurements. A systematic, self-contained introduction to
these concepts (including a simple example) are given in Appendices A-E.

\section{Adaptive sensing of magnetic noises}

\label{SEC_GENERAL}

The main purpose of this section is to identify the distinguishing features of
spin-based noise sensing compared with the estimation of deterministic
Hamiltonian parameters. For this purpose, we consider a generic pure-dephasing
model%
\begin{equation}
\hat{H}(t)=[\omega+\tilde{\omega}(t)]\hat{S}_{z}\label{HT}%
\end{equation}
describing the evolution of a spin-1/2 $\mathbf{\hat{S}}$ under a constant
magnetic field $B$ and a magnetic noise $\tilde{B}(t)$ along the $z$ axis,
where $\omega\equiv\gamma B$, $\tilde{\omega}(t)\equiv\gamma\tilde{B}(t)$, and
$\gamma$ is the gyromagnetic ratio. This model is relevant to many experiments
involving a localized electron spin in solid state environments (such as the
semiconductor quantum dots and nitrogen-vacancy centers in diamond), where the
dominant magnetic noises come from the surrounding electron spin bath or
nuclear spin bath. The former can be modelled by a Ornstein--Uhlenbeck noise
\cite{LangeScience2010,DobrovitskiPRL2009,WitzelPRB2012,WitzelPRB2014}, while
the latter can be modelled by a quasi-static noise
\cite{ShulmanNatCommun2014,DelbecqPRL2016} (see Ref. \onlinecite{YangRPP2017}
for a review). Next, we follow the standard steps outlined in Sec. II and
further detailed in Appendices A-E to discuss the estimation of the noise
$\tilde{\omega}(t)$ in comparison with the estimation of $\omega$ -- a
paradigmatic Hamiltonian parameter.

For step 1, we prepare the spin into a pure initial state
\begin{equation}
|\psi_{\mathrm{in}}\rangle=\cos\frac{\Theta}{2}|\uparrow\rangle+\sin
\frac{\Theta}{2}|\downarrow\rangle\label{PHI_IN}%
\end{equation}
parametrized by $\Theta$. Next, under the Hamiltonian in Eq. (\ref{HT}), the
spin evolves for an interval $\tau$ into a final mixed state
\begin{equation}
\hat{\rho}(\tau)=\left(
\begin{array}
[c]{cc}%
\cos^{2}\frac{\Theta}{2} & e^{-i\omega\tau}e^{-\chi}\sin\frac{\Theta}{2}%
\cos\frac{\Theta}{2}\\
e^{i\omega\tau}e^{-\chi^{\ast}}\sin\frac{\Theta}{2}\cos\frac{\Theta}{2} &
\sin^{2}\frac{\Theta}{2}%
\end{array}
\right)  ,\label{RHO_TAU}%
\end{equation}
where $e^{-\chi}\equiv\langle e^{-i\int_{0}^{\tau}\tilde{\omega}(t)dt}\rangle$
is the average of the random phase over the noise distribution. For general
noies, $\chi$ could be complex. Here we assume $\tilde{\omega}(t)$ is
symmetric about zero, then $\chi$ is real. Usually the fluctuation of the
random phase grows with the evolution time $\tau$, so $e^{-\chi}$ decreases
with $\tau$, corresponding to the decay of the average spin in the $xy$ plane
or spin decoherence for short \cite{YangRPP2017}. From Eq. (\ref{RHO_TAU}), we
see that all the information about $\omega$ is carried by the phase factor
$e^{-i\omega\tau}$, while all the information about the noise is carried by
the decoherence factor $e^{-\chi}$.

For $\omega$ and \textit{any} parameter (denoted by $\zeta$)\ that
characterizes the noise $\tilde{\omega}(t)$, the QFI in the final state
$\hat{\rho}(\tau)$ can be computed by Eq. (\ref{QFI_TLS}) as
\begin{subequations}
\label{QFI0}%
\begin{align}
\mathcal{F}_{\omega} &  =\tau^{2}e^{-2\chi}\sin^{2}\Theta,\label{QFI_W}\\
\mathcal{F}_{\zeta} &  =\frac{(\partial_{\zeta}\chi)^{2}}{e^{2\chi}-1}\sin
^{2}\Theta.\label{QFI_Z}%
\end{align}
Here $\mathcal{F}_{\omega}$ and $\mathcal{F}_{\zeta}$ show very different
dependences on the decoherence factor $\chi$. This highlights the first
distinguishing feature of noise sensing compared with the estimation of a
deterministic magnetic field. For estimating $\omega$ ($\zeta$), we should
maximize $\mathcal{F}_{\omega}$ ($\mathcal{F}_{\zeta}$) by tuning the
controlling parameters \footnote{In principle, we can also apply
time-dependent control on the spin during the evolution to engineer the final
state and then maximize $\mathcal{F}_{\omega}$ or $\mathcal{F}_{\zeta}$ in the
final state by optimizing this control. However, such optimization for a
time-dependent, open quantum system is still an open issue
\cite{YuanPRL2016,YuanPRL2015,PangNC2017}.} $\Theta$ and $\tau$. The optimal
value of $\Theta$ is $\Theta=\pi/2$. The optimal value of $\tau$ should be
chosen to maximize $\mathcal{F}_{\omega}$ ($\mathcal{F}_{\zeta})$. The optimal
$\tau$ depends on the specific form of $\chi$ as a function of $\tau$, which
in turn is determined by the details of the noise (to be discussed shortly).

For step 2, we consider a general projective measurement on the spin-1/2 along
an axis with polar angle $\theta$ and azimuth $\varphi$. This measurement on
$\hat{\rho}(\tau)$ gives an outcome $\pm1$ according to the probability
distribution%
\end{subequations}
\begin{equation}
P_{\pm1}=\frac{1\pm e^{-\chi}\sin\theta\cos(\omega\tau-\varphi)}{2}.
\label{PPM}%
\end{equation}
Here $P_{\pm1}$ as a function of $\omega$ has a period $2\pi/\tau$, thus the
measurement cannot distinguish $\omega$ and $\omega+2n\pi/\tau$ ($n\in
\mathbb{Z}$). This is the commonly encountered modulo-2$\pi$ ambiguity problem
in Hamiltonian parameter estimation. By contrast, $P_{\pm1}$ are usually not
periodic in the noise parameters, so the modulo-2$\pi$ ambiguity is absent.
This highlights the second distinguishing feature of noise sensing.

\begin{figure}[ptb]
\includegraphics[width=\columnwidth]{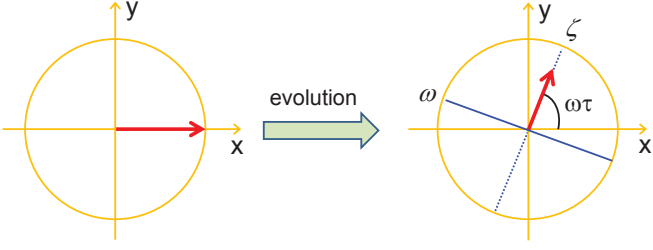}\caption{Evolution of a
spin-1/2 driven by the Hamiltonian in Eq. (\ref{HT}). The red arrows denote
the initial and final spin orientation and the solid (dashed) blue line
denotes the optimal measurement axis for estimating $\omega$ (any noise
parameter $\zeta$).}%
\label{G_EVOLVE}%
\end{figure}

Given the measurement distribution, we can compute the CFI from Eq.
(\ref{CFI_DEF}) and obtain
\begin{subequations}
\label{CFI0}%
\begin{align}
F_{\omega}  &  =\tau^{2}\frac{\sin^{2}\theta\sin^{2}(\omega\tau-\varphi
)}{e^{2\chi}-\sin^{2}\theta\cos^{2}(\omega\tau-\varphi)},\label{CFI_W}\\
F_{\zeta}  &  =(\partial_{\zeta}\chi)^{2}\frac{\sin^{2}\theta\cos^{2}%
(\omega\tau-\varphi)}{e^{2\chi}-\sin^{2}\theta\cos^{2}(\omega\tau-\varphi)}.
\label{CFI_Z}%
\end{align}
We set $\theta=\pi/2$ and $\varphi=\omega\tau+\pi/2$ ($\varphi=\omega\tau$),
so the CFI attains the QFI: $F_{\omega}=\mathcal{F}_{\omega}$ ($F_{\zeta
}=\mathcal{F}_{\zeta}$). Namely, the optimal measurement for estimating
$\omega$ ($\zeta)$ is along an axis in the $xy$ plane perpendicular (parallel)
to the spin in the final state, as shown in Fig. \ref{G_EVOLVE}. This
highlights the third distinguishing feature of noise sensing.

For step 3, we adopt the maximum likelihood estimator (see Appendix C for
details and Appendix D for an example) and leave the detailed numerical
simulation to the next section.

Finally, we discuss how to optimize the evolution time $\tau$ to maximize the
QFI $\mathcal{F}_{\omega}=\tau^{2}e^{-2\chi}$ and $\mathcal{F}_{\zeta
}=(\partial_{\zeta}\chi)^{2}/(e^{2\chi}-1)$. For any classical noise
(including static noises \cite{KropfPRX2016} and dynamical ones, Markovian
noises and non-Markovian ones \cite{KropfPRX2016,YangRPP2017}), once the
statistics of the noise $\tilde{\omega}(t)$ is given, we can determine $\chi$
and hence $\mathcal{F}_{\omega}$ and $\mathcal{F}_{\zeta}$ as functions of
$\tau$ for any classical noise, at least in principle. Thus the method
described here can be used to sensing an arbitrary classical noise, such as
the abnormal static noises due to disorder averaging \cite{KropfPRX2016}.
Moreover, although the discussions above are restricted to a single spin-1/2
(or equivalently a qubit), the method can also be used to infer the properties
of noises on a general quantum system \cite{KropfPRX2016}. Compared with the
spin-1/2 case, the difference is that the QFI should be calculated from Eq.
(\ref{QFI_ALL}) and the optimal measurement capable of converting all the QFI
contained in the final density matrix of a general quantum system into the CFI
is more complicated (see Appendix \ref{SEC_MEASUREMENT}).

Here for specificity we consider a widely used noise responsible for spin
decoherence in electron spin baths
\cite{LangeScience2010,DobrovitskiPRL2009,WitzelPRB2012,WitzelPRB2014}:\ the
Ornstein--Uhlenbeck noise, which is a Gaussian noise characterized by the
auto-correlation function
\end{subequations}
\[
\langle\tilde{\omega}(t)\tilde{\omega}(t^{\prime})\rangle=b^{2}%
e^{-|t-t^{\prime}|/\tau_{c}},
\]
with $b$ ($\tau_{\mathrm{c}}$) the amplitude (memory time) of the noise. The
Wick's theorem for Gaussian noises gives \cite{YangRPP2017}%
\begin{equation}
\chi=b^{2}\tau_{\mathrm{c}}^{2}(\frac{\tau}{\tau_{\mathrm{c}}}+e^{-\tau
/\tau_{\mathrm{c}}}-1)\approx\left\{
\begin{array}
[c]{ll}%
\dfrac{1}{2}b^{2}\tau^{2} & (\tau\ll\tau_{\mathrm{c}})\\
& \\
(b^{2}\tau_{\mathrm{c}})\tau & (\tau\gg\tau_{\mathrm{c}})
\end{array}
\right.  .\label{L}%
\end{equation}
For $\tau\ll\tau_{\mathrm{c}}$, the spin decoherence is Gaussian: $e^{-\chi
}=e^{-b^{2}\tau^{2}/2}$. For $\tau\gg\tau_{\mathrm{c}}$, the spin decoherence
is exponential $e^{-\chi}\approx e^{-\tau/T_{\varphi}}$ on a time scale
\begin{equation}
T_{\varphi}\equiv\frac{1}{b^{2}\tau_{\mathrm{c}}}.\label{TPHI}%
\end{equation}
Substituting Eq. (\ref{L}) into Eq. (\ref{QFI0}) gives the QFI's about
$\omega,b,$ and $\tau_{\mathrm{c}}$, respectively:
\begin{align*}
\mathcal{F}_{\omega} &  =\tau^{2}e^{-2\chi},\\
\mathcal{F}_{b} &  =\frac{4}{b^{2}}\frac{\chi^{2}}{e^{2\chi}-1},\\
\mathcal{F}_{\tau_{\mathrm{c}}} &  =\frac{g^{2}(\tau/\tau_{\mathrm{c}})}%
{\tau_{\mathrm{c}}^{2}}\frac{\chi^{2}}{e^{2\chi}-1},
\end{align*}
where%
\[
g(x)\equiv\frac{x-2+e^{-x}(x+2)}{x+e^{-x}-1}\approx\left\{
\begin{array}
[c]{ll}%
\frac{x}{3} & (x\ll1)\\
& \\
1 & (x\gg1)
\end{array}
\right.
\]
increases monotonically with $x$ till saturation. With increasing evolution
time $\tau$, the decoherence factor $\chi$ increases monotonically, so all the
QFI's first increases for small spin decoherence and then begin to decrease
when the spin decoherence becomes significant. For $\tau\ll\tau_{\mathrm{c}}$,
the QFI's are given by
\begin{subequations}
\label{QFI_SHORT}%
\begin{align}
\mathcal{F}_{\omega} &  =\tau^{2}e^{-b^{2}\tau^{2}},\label{FW_SHORT}\\
\mathcal{F}_{b} &  =\frac{b^{2}\tau^{4}}{e^{b^{2}\tau^{2}}-1},\label{FB_SHORT}%
\\
\mathcal{F}_{\tau_{\mathrm{c}}} &  =\frac{1}{36}\frac{b^{4}\tau^{2}}%
{e^{b^{2}\tau^{2}}-1}\left(  \frac{\tau}{\tau_{\mathrm{c}}}\right)
^{4}.\label{FTC_SHORT}%
\end{align}
For $\tau\gg\tau_{\mathrm{c}}$, the QFI's are given by
\end{subequations}
\begin{subequations}
\label{QFI_LONG}%
\begin{align}
\mathcal{F}_{\omega} &  =\tau^{2}e^{-2\tau/T_{\varphi}},\label{FW_LONG}\\
\mathcal{F}_{b} &  =4b^{2}\tau_{\mathrm{c}}^{2}\frac{\tau^{2}}{e^{2\tau
/T_{\varphi}}-1},\label{FB_LONG}\\
\mathcal{F}_{\tau_{\mathrm{c}}} &  =b^{4}\frac{\tau^{2}}{e^{2\tau/T_{\varphi}%
}-1}.\label{FTC_LONG}%
\end{align}
Next, we discuss the the optimization of the evolution time $\tau$ to maximize
the QFI.

\begin{figure}[ptb]
\includegraphics[width=\columnwidth]{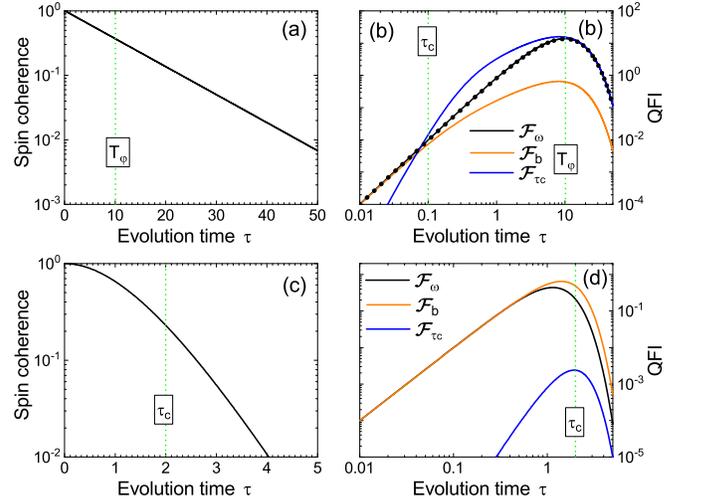}\caption{(a), (c) spin coherence
and (b), (d) quantum Fisher information as functions of the evolution time
$\tau$. (a) and (b) for Markovian noise $b=1$, $\tau_{\mathrm{c}}=0.1$; (c)
and (d) for non-Markovian noise $b=1,$ $\tau_{\mathrm{c}}=2$. The black dotted
line in (b) for Eq. (\ref{FW_LONG}).}%
\label{G_QFI}%
\end{figure}

For Markovian noises (i.e., $b\tau_{\mathrm{c}}\ll1$), the spin coherence
decays exponentially $e^{-\chi}\approx e^{-\tau/T_{\varphi}}$ on a time scale
$T_{\varphi}$, as shown in Fig. \ref{G_QFI}(a). Thus $\mathcal{F}_{\omega}$ is
well approximated by Eq. (\ref{FW_LONG}), shown as the black dotted line in
Fig. \ref{G_QFI}(b). With increasing evolution time $\tau$, $\mathcal{F}%
_{\omega}$ first increases quadratically and then decays exponentially. At the
optimal evolution time $\tau_{\omega,\mathrm{opt}}=T_{\varphi},$ it reaches
the maximum $\mathcal{F}_{\omega,\mathrm{opt}}=T_{\varphi}^{2}/e^{2}$, as
shown in Fig. \ref{G_QFI}(b). For noise sensing, $\mathcal{F}_{b}$ and
$\mathcal{F}_{\tau_{\mathrm{c}}}$ as functions of $\tau$ differ from that of
$\mathcal{F}_{\omega}$ in that they exhibit three stages [Fig. \ref{G_QFI}%
(b)]. For $\tau\ll\tau_{\mathrm{c}}$, we have $\mathcal{F}_{b}\approx\tau^{2}$
and $\mathcal{F}_{\tau_{\mathrm{c}}}\propto\tau^{4}$. For $\tau_{\mathrm{c}%
}\ll\tau\ll T_{\varphi}$, $\mathcal{F}_{b}$ and $\mathcal{F}_{\tau
_{\mathrm{c}}}$ increase linearly with $\tau$. For $\tau\gg T_{\varphi}$,
$\mathcal{F}_{b}$ and $\mathcal{F}_{\tau_{\mathrm{c}}}$ decays exponentially
with $\tau$. At the optimal evolution time $\tau_{b,\mathrm{opt}}\approx
\tau_{\tau_{\mathrm{c}},\mathrm{opt}}\approx0.8T_{\varphi}$, $\mathcal{F}_{b}$
and $\mathcal{F}_{\tau_{\mathrm{c}}}$ attain their maxima $\mathcal{F}%
_{b,\mathrm{opt}}\approx0.65/b^{2}$ and $\mathcal{F}_{\tau_{\mathrm{c}%
},\mathrm{opt}}\approx0.162/\tau_{\mathrm{c}}^{2}$. For estimating $\omega$,
the optimal evolution time $\tau_{\omega,\mathrm{opt}}$ is independent of
$\omega$, thus adaptive measurements are not necessary. By contrast, for
estimating the noise parameter $b$ ($\tau_{\mathrm{c}}$), the optimal
evolution time $\tau_{b,\mathrm{opt}}$ ($\tau_{\tau_{\mathrm{c}},\mathrm{opt}%
}$)\ depend on the parameter $b$ ($\tau_{\mathrm{c}}$) to be estimated, so
adaptive measurements are crucial.

For non-Markovian noises (i.e., $b\tau_{\mathrm{c}}\gtrsim1$), the Gaussian
decay $e^{-\chi}=e^{-b^{2}\tau^{2}/2}$ for the spin coherence and the QFI's
[Eq. (\ref{QFI_SHORT})] becomes appreciable even in the short-time regime
$\tau\ll\tau_{\mathrm{c}}$, as shown in Fig. \ref{G_QFI}(c) and (d). In
general, the peak location of the QFI as a function of $\tau$ depends on both
$\tau_{\mathrm{c}}$ and $b$, thus adaptive measurements are crucial for
estimating $b$ and $\tau_{\mathrm{c}}$, as opposed to the estimation of
$\omega$.

The general analysis in this section have identified a series of
distinguishing features for the sensing of noises compared to the estimation
of deterministic magnetic fields, including the different dependences of the
QFI's on the spin decoherence, the absence of the modulo-2$\pi$ phase
ambiguity, the different optimal measurement schemes, and the crucial role of
adaptive measurements. In the next section, we perform numerical simulations
to demonstrate the feasibility of adaptive measurements to improve the
precision of noise sensing.

\section{Adaptive measurement of spin decoherence time}

Here we consider the decoherence of a localized spin caused by the surrounding
electron spin bath in the solid state environment. The electron spin bath can
be modelled by a Markovian noise \cite{YangRPP2017} and leads to exponential
spin decoherence $e^{-\chi}\approx e^{-\tau/T_{\varphi}}$ on a time scale
$T_{\varphi}$ [see Eq. (\ref{TPHI})]. The spin decoherence time $T_{\varphi}$
is a key characteristics of spin qubits in solid state environments. Next, we
consider the adaptive estimation of the spin decoherence time $T_{\varphi}$.
The rapid estimation of the spin decoherence time is not only important for
the experimental characterization of the spin decoherence, but also allows us
to design efficient coherence protection schemes to suppress the decoherence.
Since $T_{\varphi}$ is a noise parameter, its adaptive estimation differs
significantly from the estimation of a Hamiltonian parameter.

\begin{figure}[ptb]
\includegraphics[width=\columnwidth]{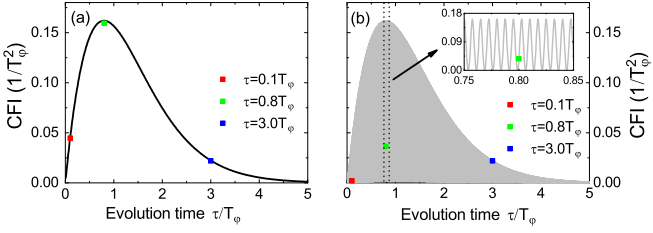}\caption{Classical Fisher
information as a function of the evolution time for (a) spin echo protocol and
(b) free evolution protocol with $\omega T_{\varphi}=400\pi/3\gg1$. The
symbols denote $(\delta T_{\varphi})^{-2}/N$, where $\delta T_{\varphi}$ is
the sensing precision from numerically simulating $N=10^{4}$ repeated
measurements with different evolution time:\ red for $\tau=0.1T_{\varphi}$,
green for $\tau=0.8T_{\varphi}$, and blue for $\tau=3T_{\varphi}$. The inset
of (b) zooms in on dashed square region.}%
\label{G_CFI}%
\end{figure}

According to Sec. \ref{SEC_GENERAL}, the optimal initial state of the spin is
Eq. (\ref{PHI_IN}) with $\Theta=\pi/2$. In the following we consider two
different protocols to measure the spin decoherence time $T_{\varphi}$: (i)
Spin echo; (ii) Free evolution.

For spin echo, we first let the spin evolve under the Hamiltonian in Eq.
(\ref{HT}) for an interval $\tau/2$, next apply an instantaneous $\pi$-pulse
to induce the flip between $|\uparrow\rangle$ and $|\downarrow\rangle$, and
finally let the spin evolve under the Hamiltonian in Eq. (\ref{HT}) for
another interval $\tau/2$ into the final state $\hat{\rho}(\tau)=(1+e^{-\tau
/T_{\varphi}}\hat{\sigma}_{x})/2$. The QFI about $T_{\varphi}$ in this final
state is
\end{subequations}
\begin{equation}
\mathcal{F}(\tau)=\frac{\tau^{2}}{T_{\varphi}^{4}}\frac{1}{e^{2\tau
/T_{\varphi}}-1}, \label{QFI_ALL}%
\end{equation}
where we have made its dependence on the evolution time $\tau$ explicit. This
spin echo technique \cite{HahnPR1950} can eliminate quasi-static noises (such
as those from the surrounding nuclear spins of the host lattice) and single
out the decoherence caused by the Markovian noise under consideration.
Interestingly, it also eliminates the Larmor frequency $\omega$, so that the
final density matrix $\hat{\rho}(\tau)$ is independent of $\omega$. Finally,
we perform a projective measurement on the spin along the $x$ axis. According
to Sec. \ref{SEC_GENERAL}, this measurement is optimal. Indeed, it gives an
outcome $\pm1$ according to the probability distribution%
\begin{equation}
P(\pm1|T_{\varphi})=\frac{1\pm e^{-\tau/T_{\varphi}}}{2}, \label{PPM_ECHO}%
\end{equation}
and the CFI in each outcome attains the QFI:
\begin{equation}
F(\tau)=\frac{\tau^{2}}{T_{\varphi}^{4}}\frac{1}{e^{2\tau/T_{\varphi}}-1},
\label{CFI_ECHO}%
\end{equation}
as shown in Fig. \ref{G_CFI}(a).

For free evolution, we simply let the spin evolve under the Hamiltonian in Eq.
(\ref{HT}) for an interval $\tau$ into the final state $\hat{\rho}%
(\tau)=1/2+(e^{-\tau/T_{\varphi}}/2)[\hat{\sigma}_{x}\cos(\omega\tau
)+\hat{\sigma}_{y}\sin(\omega\tau)]$, i.e., Eq. (\ref{RHO_TAU}) with
$\Theta=\pi/2$ and $\chi=\tau/T_{\varphi}$. This final state differs from that
of the spin echo protocol in that it still depends sensitively on the Larmor
frequency $\omega$. Consequently, in order to measure $T_{\varphi}$ from this
free-evolution final state, precise knowledge about $\omega$ is usually
necessary (to be discussed shortly), although the QFI about $T_{\varphi}$
contained in this final state is still given by Eq. (\ref{QFI_ALL}), i.e., the
same as the spin echo protocol. Finally, we should perform an optimal
measurement to convert all the QFI\ into the CFI. According to Sec.
\ref{SEC_GENERAL}, the optimal measurement is a projective one along the
azimuth $\omega\tau$ in the $xy$ plane (dashed blue line in Fig.
\ref{G_EVOLVE}), whose CFI is equal to the QFI in Eq. (\ref{QFI_ALL}).
However, in our adaptive measurement scheme (to be discussed shortly), the
parameter $\tau$ and hence the measurement axis will vary in different
measurement cycles. The frequent change of the measurement axis may
complicates its experimental realization. To avoid this problem, we fix the
measurement axis to be along the $x$ axis (i.e., we always measure
$\hat{\sigma}_{x}$), then the measurement distribution is%
\begin{equation}
P(\pm1|T_{\varphi})=\frac{1\pm e^{-\tau/T_{\varphi}}\cos(\omega\tau)}{2}
\label{PPM_FREE}%
\end{equation}
and the CFI\ in each outcome,%
\begin{equation}
F(\tau)=\frac{\tau^{2}}{T_{\varphi}^{4}}\frac{\cos^{2}(\omega\tau)}%
{e^{2\tau/T_{\varphi}}-\cos^{2}(\omega\tau)}, \label{CFI_FREE}%
\end{equation}
shows rapidly oscillation as a function of $\tau$, with its envelope
coinciding with the QFI [see Fig. \ref{G_CFI}(b)]. In other words, the CFI
still attains the QFI when $\tau$ is an integer multiple of $\pi/\omega$, but
does not attains the QFI for general $\tau$. Fortunately, we can still tune
$\tau$ to maximize the QFI and the CFI simultaneously.

Now we optimize the evolution time $\tau$. For spin echo, the optimal $\tau$
is [see Fig. \ref{G_CFI}(a)]%
\begin{equation}
\tau_{\mathrm{opt}}=0.8T_{\varphi}. \label{TOPT_ECHO}%
\end{equation}
For free evolution, under the realistic assumption $\omega\gg1/T_{\varphi}$,
the optimal $\tau$ is an integer multiple of $\pi/\omega$ closest to
$0.8T_{\varphi}$ [see Fig. \ref{G_CFI}(b)]:
\begin{equation}
\tau_{\mathrm{opt}}=n\frac{\pi}{\omega}\ (n\in\mathbb{Z})\ \mathrm{and}%
\ \tau_{\mathrm{opt}}\approx0.8T_{\varphi}. \label{TOPT_FREE}%
\end{equation}
For both the spin echo protocol and the free evolution protocol, choosing
$\tau=\tau_{\mathrm{opt}}$ gives the same maximal CFI and QFI,
\begin{equation}
F_{\mathrm{opt}}=\mathcal{F}_{\mathrm{opt}}=\frac{0.16}{T_{\varphi}^{2}},
\label{FOPT}%
\end{equation}
and hence the same optimal sensing precision%
\begin{equation}
(\delta T_{\varphi})_{\mathrm{opt}}=\frac{2.5T_{\varphi}}{\sqrt{N}}
\label{DT_OPT}%
\end{equation}
for $N$ repeated measurements. The difference is that in the spin echo (free
evolution) protocol, $\tau_{\mathrm{opt}}$ is independent of (dependent on)
the Larmor frequency $\omega$. Specifically, in the free evolution protocol,
the rapid oscillation of the CFI as a function of $\tau$ with a period
$\pi/\omega$ [see Fig. \ref{G_CFI}(b)] requires precise knowledge about
$\omega$ and high control precision of $\tau$ on the order of $1/\omega$ to
correctly locate the maximum [i.e., Eq. (\ref{TOPT_FREE})] of the CFI. By
contrast, in the spin echo protocol, the CFI as a function of $\tau$ is
independent of $\omega$ [see Fig. \ref{G_CFI}(a)], so it requires \textit{no}
knowledge about $\omega$ and relatively low control precision of $\tau$ on the
order of $1/T_{\varphi}$ ($\gg1/\omega$) to correctly locate the maximum of
the CFI.

Unfortunately, for both protocols, $\tau_{\mathrm{opt}}$ depends on the
unknown parameter $T_{\varphi}$. Due to this dependence, adaptive schemes that
update $\tau_{\mathrm{opt}}$ after each measurement cycle can outperform
significantly non-adaptive ones. For each protocol, we consider three
different measurement schemes involving different treatments of the evolution
time $\tau$: repeated measurements, adaptive measurements, and the
least-square fitting that is commonly used in experiments.

\subsection{Repeated measurement scheme}

The evolution time $\tau$ is fixed during the entire estimation process. After
repeating the initialization-evolution-measurement cycle $N$ times, we get $N$
outcomes $\mathbf{u}=(u_{1},u_{2},\cdots,u_{N})$. Using these outcomes, we
refine our knowledge about $T_{\varphi}$ to the posterior distribution
\[
P_{\mathbf{u}}(T_{\varphi})\sim\lbrack P(+1|T_{\varphi})]^{N_{+}%
}[P(-1|T_{\varphi})]^{N_{-}},
\]
where $N_{+}$ ($N_{-}$) is the number of outcome $+1$ ($-1$) and
$P(\pm1|T_{\varphi})$ is given by Eq. (\ref{PPM_ECHO}) for spin echo and Eq.
(\ref{PPM_FREE}) for free evolution. Finally, we construct the maximum
likelihood estimator $T_{\mathrm{M}}\equiv\arg\max P_{\mathbf{u}}(T_{\varphi
})$ and quantify its precision by [cf. Eq. (\ref{DZ_DEF})]
\begin{equation}
\delta T_{\varphi}\equiv\sqrt{\int(T_{\mathrm{M}}-T_{\varphi})^{2}%
P_{\mathbf{u}}(T_{\varphi})dT_{\varphi}}. \label{DT_DEF}%
\end{equation}

To analyze the performance of this scheme, we notice that for large $N$, the
maximum likelihood estimator is known to be unbiased and can saturate the
Cram\'{e}r-Rao bound Eq. (\ref{CRB}), so the sensing precision can be
approximated by%
\begin{equation}
(\delta T_{\varphi})_{\mathrm{CRB}}\equiv\frac{1}{\sqrt{NF(\tau)}}.
\label{DT_CRB}%
\end{equation}
Here the CFI$\ F(\tau)$ is given by Eq. (\ref{CFI_ECHO}) for spin echo and Eq.
(\ref{CFI_FREE}) for free evolution (see Fig. \ref{G_CFI}). The evolution time
$\tau$ directly determines the sensing precision, e.g., setting $\tau
=\tau_{\mathrm{opt}}$ would lead to the optimal sensing precision in Eq.
(\ref{DT_OPT}). However, $\tau_{\mathrm{opt}}$ is \textit{unknown} because it
depends on the unknown parameter $T_{\varphi}$ to be estimated. This makes
adaptive measurements crucial for achieving the optimal sensing precision.

\subsection{Adaptive measurement schemes}

The key idea is to use the outcomes of previous measurement to refine our
knowledge about $T_{\varphi}$ and then use this knowledge to optimize $\tau$.
We consider two different adaptive schemes:\ the CFI-based scheme
\cite{OlivaresJPB2009,BrivioPRA2010,PangNC2017} and the locally optimal
adaptive scheme \cite{BerryPRL2000,BerryPRA2001,SaidPRB2011,SergeevichPRA2011}%
, as introduced in Appendix E. The former updates the maximum likelihood
estimator $T_{\mathrm{M}}$ after every initialization-evolution-measurement
cycle and then set the evolution time to
\begin{equation}
\tau=0.8T_{\mathrm{M}}\label{TAU_ECHO}%
\end{equation}
for the spin echo protocol or
\begin{equation}
\tau=n\frac{\pi}{\omega}\ \mathrm{and}\ \tau\approx0.8T_{\mathrm{M}}%
\ (n\in\mathbb{Z})\label{TAU_FREE}%
\end{equation}
for the free evolution protocol. The latter optimizes $\tau$ to minimize the
expected uncertainty of the estimator at the end of the next cycle (see
Appendix E). Suppose at the end of the $(n-1)$th cycle, our knowledge about
$T_{\varphi}$ is quantified by the distribution $P(T_{\varphi})$ and the
maximum likelihood estimator $T_{\mathrm{M}}\equiv\arg\max P(T_{\varphi})$
constructed from the outcomes of all the previous cycles. In the $n$th cycle
with the evolution time $\tau$, the measurement distribution $P(\pm
1|T_{\varphi})$ [Eq. (\ref{PPM_ECHO}) or Eq. (\ref{PPM_FREE})] depends on
$T_{\varphi}$ and $\tau$. If the outcome is $u$, then our knowledge would be
refined to $P_{u}(T_{\varphi})\sim P(T_{\varphi})P(u|T_{\varphi})$, which in
turn gives the maximum likelihood estimator $T_{\mathrm{M}}(u,\tau)\equiv
\arg\max P_{u}(T_{\varphi})$ and its uncertainty [cf. Eq. (\ref{DZ_DEF})]
\[
\delta T_{\varphi}(u,\tau)\equiv\sqrt{\int[T_{\varphi}-T_{\mathrm{M}}%
(u,\tau)]^{2}P_{u}(T_{\varphi})dT_{\varphi}}.
\]
Since the probability for outcome $u$ is estimated as $P(u|T_{\mathrm{M}})$,
we should choose $\tau$ in the $n$th cycle to minimize the expected
uncertainty [cf. Eq. (\ref{BERRY})]\
\[
\overline{\delta T_{\varphi}}(\tau)\equiv\sum_{u=\pm1}P(u|T_{\mathrm{M}%
})\delta T(u,\tau).
\]
For $n=1$, i.e., the first cycle, there is no prior information, i.e.,
$P(T_{\varphi})$ is a constant, so $T_{\mathrm{M}}$ is chosen randomly.

To analyze the performance, we notice that after a large number of adaptive
steps, the estimator $T_{\mathrm{M}}$ would approach the true decoherence time
$T_{\varphi}$. Consequently, according to Appendix E, the evolution time
$\tau$ and hence the sensing precision for these two adaptive schemes would
coincide with each other. In addition, the evolution time $\tau$ in Eqs.
(\ref{TAU_ECHO}) and (\ref{TAU_FREE}) would approach $\tau_{\mathrm{opt}}$, so
the corresponding sensing precision would approach the optimal precision
$(\delta T_{\varphi})_{\mathrm{opt}}$ in Eq. (\ref{DT_OPT}).

\subsection{Least-square fitting scheme}

For a given range $[0,\tau_{\max}]$ of the evolution time, we uniformly
discretize it into $M\gg1$ grids $\tau_{k}=k\Delta\tau$, where $\Delta
\tau=\tau_{\max}/M$. For each $\tau_{k}$, we repeat the $\hat{\sigma}_{x}$
measurement $\nu$ times and calculate their average. Then we fit this average
as a function of $\tau$ to the theoretical curve $\langle\hat{\sigma}%
_{x}\rangle=e^{-\tau/T_{\varphi}}$ (for the spin echo protocol) or
$\langle\hat{\sigma}_{x}\rangle=\cos(\omega\tau)e^{-\tau/T_{\varphi}}$ (for
the free evolution protocol) to obtain an estimator to $T_{\varphi}$. Finally,
we repeat the procedures above for $q\gg1$ times to obtain many estimators
$T_{\varphi,\mathrm{est}}^{(1)},T_{\varphi,\mathrm{est}}^{(2)},\cdots
T_{\varphi,\mathrm{est}}^{(q)}$ and determine the uncertainty $\delta
T_{\varphi}$ of a single estimator as the square root of the statistical
variance these estimators, i.e.,
\[
\delta T_{\varphi}\equiv\frac{1}{q}\sum_{i=1}^{q}(T_{\varphi,\mathrm{est}%
}^{(i)}-\bar{T}_{\varphi,\mathrm{est}})^{2},
\]
where $\bar{T}_{\varphi,\mathrm{est}}\equiv(1/q)\sum_{i=1}^{q}T_{\varphi
,\mathrm{est}}^{(i)}$.

According to the Cram\'{e}r-Rao bound in Eq. (\ref{CRB}), the sensing
precision of this scheme can be roughly estimated as%
\[
(\delta T_{\varphi})_{\mathrm{CRB}}\equiv\frac{1}{\sqrt{NF_{\mathrm{ave}}}},
\]
where $N\equiv\nu M$ is the total number of measurements,
\[
F_{\mathrm{ave}}=\frac{1}{\tau_{\max}}\int_{0}^{\tau_{\max}}F(\tau)d\tau
\]
is the average of the CFI over the range $[0,\tau_{\max}]$. For spin echo,
$F(\tau)$ in Eq. (\ref{CFI_ECHO}) decays exponentially for large $\tau$ [see
Fig. \ref{G_CFI}(a)]. For $\tau_{\max}\gg T_{\varphi}$, we have
$F_{\mathrm{ave}}\approx0.3/(T_{\varphi}\tau_{\max})$, thus
\begin{equation}
(\delta T_{\varphi})_{\mathrm{CRB}}\approx1.8\sqrt{\frac{T_{\varphi}\tau
_{\max}}{N}}\approx0.7\sqrt{\frac{\tau_{\max}}{T_{\varphi}}}(\delta
T_{\varphi})_{\mathrm{opt}}\label{DT_CRB_AVE}%
\end{equation}
degrades monotonically with increasing $\tau_{\max}$. For free evolution,
$F(\tau)$ in Eq. (\ref{CFI_FREE}) shows rapid oscillations as a function of
$\tau$ with an envelope coinciding with the CFI for spin echo [see Fig.
\ref{G_CFI}(b)]. For $\tau_{\max}\gg T_{\varphi}$, we have $F_{\mathrm{ave}%
}\approx0.14/(T_{\varphi}\tau_{\max})$, which is about half that of the spin
echo, thus $(\delta T_{\varphi})_{\mathrm{CRB}}\approx\sqrt{\tau_{\max
}/T_{\varphi}}(\delta T_{\varphi})_{\mathrm{opt}}$ also degrades with
increasing $\tau_{\max}$.

\subsection{Numerical simulations}

\begin{figure}[ptb]
\includegraphics[width=\columnwidth]{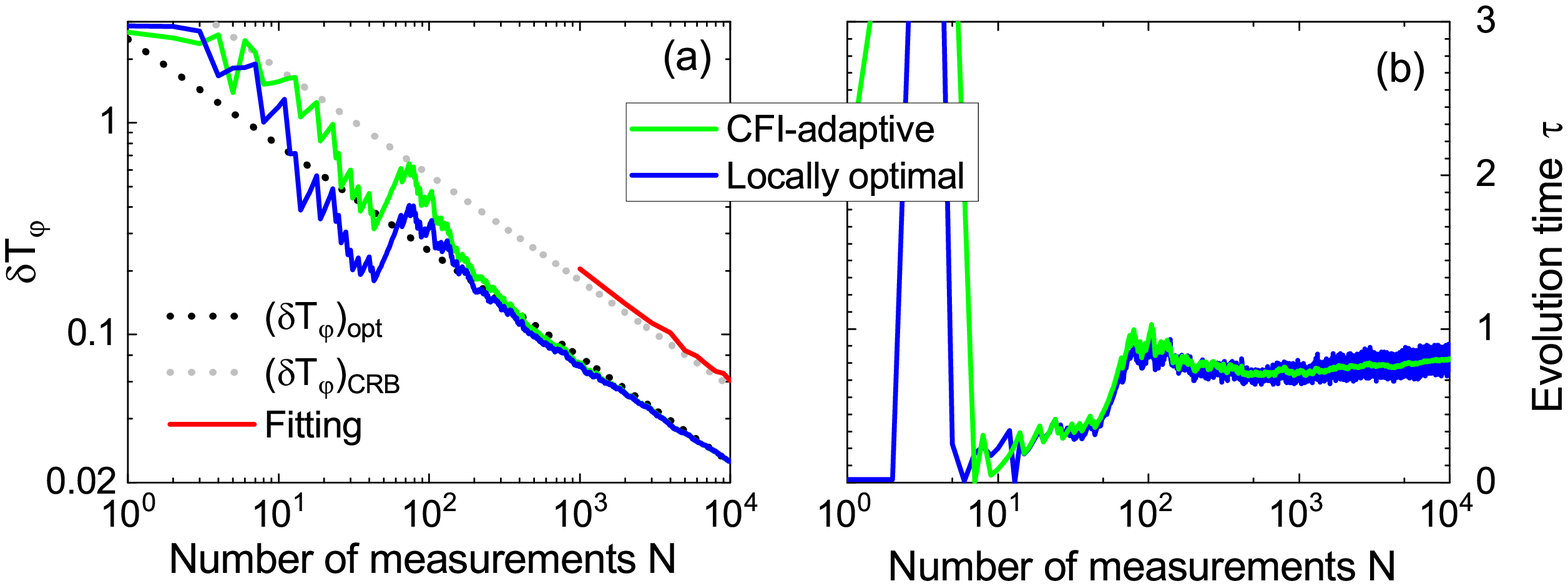}\caption{Numerical
simulation for the estimation of spin decoherence time $T_{\varphi}$ by spin
echo. (a) Estimation precision $\delta T_{\varphi}$ by least-square fitting
(red line) with $\tau_{\max}=10T_{\varphi}$ and $\nu=q=100$, CFI-based
adaptive sensing (green line) and locally optimal adaptive sensing (blue
line). The black (gray)\ dotted line indicates $(\delta T_{\varphi
})_{\mathrm{opt}}$ in Eq. (\ref{DT_OPT}) [$(\delta T_{\varphi})_{\mathrm{CRB}%
}$ in Eq. (\ref{DT_CRB_AVE})]. (b) Successive refinement of the evolution time
$\tau$ in CFI-based (green line) and locally optimal (blue line) adaptive
sensing.}%
\label{G_ECHO}%
\end{figure}

In the above, we have presented two \textit{protocols} to measure the spin
decoherence time $T_{\varphi}$: the spin echo protocol and the free evolution
protocol. For each protocol, we consider three kinds of \textit{schemes},
which involve different treatments of the evolution time $\tau$: (i) The
repeated measurement scheme uses a fixed $\tau$; (ii) The two adaptive
measurement schemes update $\tau$ in every measurement cycle; (iii) The
least-square fitting scheme scan $\tau$ over a fixed range. The spin echo
protocol singles out the $T_{\varphi}$ process from all the other unwanted
evolution, so the measurement distribution $P(\pm1|T_{\varphi})$ is
independent of the Larmor frequency $\omega$. Consequently, all the schemes
applied to the spin echo protocol require no knowledge about $\omega$ and a
relatively low control precision (on the order of $1/T_{\varphi})$ over $\tau
$. By contrast, the free evolution protocol leaves the Larmor precession
intact, so the measurement distribution $P(\pm1|T_{\varphi})$ [see Eq.
(\ref{PPM_FREE})] depends sensitively on $\omega$. Consequently, all the
schemes applied to this protocol require precise knowledge about $\omega$ and
much higher (on the order $1/\omega)$ control precision over $\tau$.
Specifically: (i)\ In the repeated measurement scheme, since the posterior
distribution $P_{\mathbf{u}}(T_{\varphi})$ depends on $\omega$, we cannot find
the maximum likelihood estimator $T_{\mathrm{M}}\equiv\arg\max P_{\mathbf{u}%
}(T_{\varphi})$ if $\omega$ is unknown; (ii)\ In the CFI-based adaptive
scheme, we cannot set $\tau$ to Eq. (\ref{TAU_FREE}) if $\omega$ is unknown;
in the locally optimal adaptive scheme, the expected uncertainty
$\overline{\delta T_{\varphi}}(\tau)$ depend on both $\tau$ and $\omega$, so
we cannot find the minimum of $\overline{\delta T_{\varphi}}(\tau)$ as a
function of $\tau$ if $\omega$ is unknown. (iii) In the least-square fitting
scheme, it would be difficult to choose the grid spacing $\Delta\tau$ and to
fit the measurement data to $\langle\hat{\sigma}_{x}\rangle=\cos(\omega
\tau)e^{-\tau/T_{\varphi}}$ to extract $T_{\varphi}$ when $\omega$ is unknown.
Therefore, the spin echo protocol is advantageous over the free evolution
protocol if our knowledge about $\omega$ is limited or the available control
precision over $\tau$ is low.

In all our numerical simulations, we take the true value $T_{\varphi}$ as the
unit of time, i.e., $T_{\varphi}=1$, and take the true value of $\omega$ to be
$\omega=400\pi/3\gg1/T_{\varphi}$.

To begin with, we check the sensing precision of the repeated measurement
scheme applied to the spin echo protocol and the free evolution protocol. We
consider three sets of evolution time: $\tau=0.1,$ $0.8$, and $3$. For each
case, our numerical simulations show that with increasing number $N$ of
repeated measurements, the uncertainty $\delta T_{\varphi}$ calculated from
Eq. (\ref{DT_DEF}) gradually approaches the large-$N$ limit $(\delta
T_{\varphi})_{\mathrm{CRB}}$ [Eq. (\ref{DT_CRB})] for both protocols. For
example, the uncertainty $\delta T_{\varphi}$ for $N=10^{4}$ repeated
measurements agree well with $(\delta T_{\varphi})_{\mathrm{CRB}}$, i.e.,
$(\delta T_{\varphi})^{-2}/N$ agree well with the CFI $F(\tau)$, as shown in
Fig. \ref{G_CFI}(a) and \ref{G_CFI}(b).

Next, we consider the spin echo protocol and compare the adaptive scheme with
the commonly used least-square fitting scheme with $\tau_{\max}=10$ and
$\nu=q=100$. As shown in Fig. \ref{G_ECHO}(a), the precision of the
least-square fitting is well approximated by $(\delta T_{\varphi
})_{\mathrm{CRB}}$ in Eq. (\ref{DT_CRB_AVE}), which is significantly worse
than the optimal sensing precision $(\delta T_{\varphi})_{\mathrm{opt}}$ in
Eq. (\ref{DT_OPT}). By contrast, the precision of both the CFI-based adaptive
scheme and the locally optimal adaptive scheme approaches the optimal
precision $(\delta T_{\varphi})_{\mathrm{opt}}$ after $\sim100$ measurements.
Physically, this is because both adaptive schemes successively adjust the
evolution time [e.g., Eqs. (\ref{TAU_ECHO}) and (\ref{TAU_FREE}) for the
CFI-based adaptive scheme applied to the spin echo protocol and the free
evolution protocol] based on the newest knowledge about the unknown parameter
$T_{\varphi}$ after every measurement. As shown in Fig. \ref{G_ECHO}(b), after
$\sim100$ measurements, the evolution time in both adaptive schemes already
approaches the optimal evolution time $\tau_{\mathrm{opt}}$ [Eq.
(\ref{TOPT_ECHO})].

\begin{figure}[ptb]
\includegraphics[width=\columnwidth]{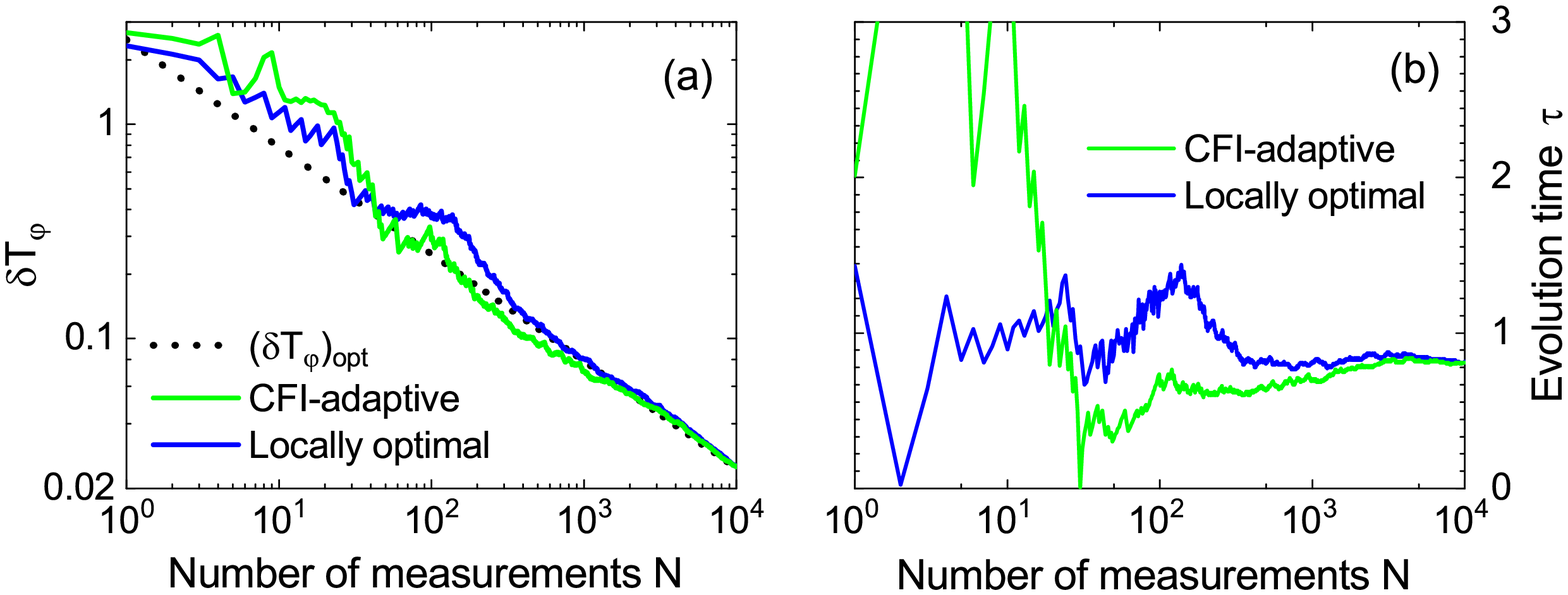}\caption{Numerical
simulation for the estimation of the spin decoherence time $T_{\varphi}$ by
free evolution. (a) Estimation precision $\delta T_{\varphi}$ by CFI-based
adaptive sensing (green line) and locally optimal adaptive sensing (blue
line). The black dotted line indicates the optimal sensing precision $(\delta
T_{\varphi})_{\mathrm{opt}}$ in Eq. (\ref{DT_OPT}). (b) Successive refinement
of the evolution time $\tau$ in CFI-based (green lines) and locally optimal
(blue lines) adaptive sensing.}%
\label{G_FREE}%
\end{figure}

Then, we turn to the free evolution protocol, which requires precise knowledge
about $\omega$ and precise control over $\tau$ on the order $1/\omega$. Since
the average spin $\langle\hat{\sigma}_{x}\rangle=\cos(\omega\tau
)e^{-\tau/T_{\varphi}}$ exhibits rapid oscillations as functions of the
evolution time $\tau$, the best way to do least-square fitting is to let the
grid spacing $\Delta\tau$ be an integer multiple of $\pi/\omega$, so that
$\tau_{k}=k\Delta\tau$ samples the envelope of the $\langle\hat{\sigma}%
_{x}\rangle$ curve only. This also amounts to sampling the envelope of the
rapidly oscillating CFI in Eq. (\ref{CFI_FREE}), so that $F(\tau
_{k})=\mathcal{F}(\tau_{k})$ attains the corresponding QFI. Therefore, the
least-square fitting scheme applied to the free evolution protocol would give
the same precision as it does for the spin echo protocol, so we do not
simulate this case any more. For the two adaptive schemes applied to the free
evolution protocol, as shown in Fig. \ref{G_FREE}(a), both the CFI-based one
and the locally optimal one approach the optimal sensing precision $(\delta
T_{\varphi})_{\mathrm{opt}}$ after $\sim100$ measurements, similar to the case
of the spin echo protocol. As shown in Fig. \ref{G_FREE}(b), the adaptive
schemes successively refine the evolution time [e.g., Eq. (\ref{TAU_FREE})]
based on the newest knowledge about $T_{\varphi}$ after every measurement.
After $\sim100$ measurements, the evolution time in both adaptive schemes
approach the optimal evolution time $\tau_{\mathrm{opt}}$ [Eq.
(\ref{TOPT_FREE})].

\section{Conclusion}

Using localized spins as ultrasensitive quantum sensors is attracting
widespread interest. Recently, adaptive measurements were used to improve the
dynamic range for the spin-based estimation of \textit{deterministic}
Hamiltonian parameters such as the external magnetic field. Here we explore a
very different direction -- the use of adaptive measurements in spin-based
sensing of \textit{random} noises. We have performed general analysis that
identifies a series of important differences between noise sensing and the
estimation of deterministic magnetic fields, such as the different dependences
on the spin decoherence, the different optimal measurement schemes, the
absence of the modulo-2$\pi$ phase ambiguity, and the crucial role of adaptive
measurement. We have also performed numerical simulations that clearly
demonstrate significant speed up of the characterization of the spin
decoherence time via adaptive measurements compared with the commonly used
least-square fitting method. This work paves the way towards adaptive noise sensing.

\begin{acknowledgments}
This work was supported by the MOST of China (Grants No. 2014CB848700), the
National Key R\&D Program of China (Grants No. 2017YFA0303400), the NSFC
(Grants No. 11774021), and the NSFC program for \textquotedblleft Scientific
Research Center\textquotedblright\ (Grant No. U1530401). We acknowledge the
computational support from the Beijing Computational Science Research Center (CSRC).
\end{acknowledgments}

\appendix{}

\section{State preparation and encoding:\ quantum Fisher information}

The amount of information about $\zeta$ contained in a general $\zeta
$-dependent quantum state $\hat{\rho}_{\zeta}$ is quantified by its QFI
\cite{BraunsteinPRL1994}%
\begin{equation}
\mathcal{F}\equiv\operatorname*{Tr}\hat{\rho}_{\zeta}\hat{L}_{\zeta}%
^{2},\label{QFI_DEF}%
\end{equation}
where $\hat{L}_{\zeta}$ is the so-called symmetric logarithmic derivative
operator: it is an Hermitian operator defined through \cite{HelstromBook1976}
\[
\partial_{\zeta}\hat{\rho}_{\zeta}=\frac{1}{2}(\hat{L}_{\zeta}\hat{\rho
}_{\zeta}+\hat{\rho}_{\zeta}\hat{L}_{\zeta}).
\]
The QFI defined in Eq. (\ref{QFI_DEF}) remains invariant under any $\zeta
$-independent unitary transformations, i.e., such transformations conserves
the quantum information. For a pure state $\hat{\rho}_{\zeta}=|\Phi
\rangle\langle\Phi|$, we have $\hat{L}_{\zeta}=2\partial_{\zeta}\hat{\rho
}_{\zeta}$ and hence
\begin{equation}
\mathcal{F}=4(\langle\partial_{\zeta}\Phi|\partial_{\zeta}\Phi\rangle
-|\langle\Phi|\partial_{\zeta}\Phi\rangle|^{2})\longrightarrow4G_{\mathrm{rms}%
}^{2},\label{QFI_PURE}%
\end{equation}
where the last step applies to unitary evolution $|\Phi\rangle=e^{-i\zeta
\hat{G}}|\Phi_{\mathrm{in}}\rangle$ and $G_{\mathrm{rms}}\equiv(\langle
\Phi_{\mathrm{in}}|\hat{G}^{2}|\Phi_{\mathrm{in}}\rangle-\langle
\Phi_{\mathrm{in}}|\hat{G}|\Phi_{\mathrm{in}}\rangle^{2})^{1/2}$ is the
root-mean-square fluctuation of $\hat{G}$ in the initial state. For a general
mixed state with the spectral decomposition $\hat{\rho}_{\zeta}=\sum_{n}%
p_{n}|\Phi_{n}\rangle\langle\Phi_{n}|$, its QFI is
\cite{KnyshPRA2011,ZhangPRA2013,LiuPRA2013}%
\begin{equation}
\mathcal{F}=\sum_{n}\frac{(\partial_{\zeta}p_{n})^{2}}{p_{n}}+\sum_{n}%
p_{n}\mathcal{F}_{n}-\sum_{m\neq n}\frac{8p_{m}p_{n}}{p_{m}+p_{n}}\left\vert
\left\langle \Phi_{m}\right\vert \partial_{\zeta}\Phi_{n}\rangle\right\vert
^{2},\label{QFI_GENERAL}%
\end{equation}
where $\{p_{n}\}$ are \textit{nonzero} eigenvalues of $\hat{\rho}_{\zeta}$,
$\{|\Phi_{n}\rangle\}$ are the corresponding ortho-normalized eigenstates, and
$\mathcal{F}_{n}$ is the QFI of the pure state $|\Phi_{n}\rangle$ [see Eq.
(\ref{QFI_PURE})]. This expression shows that the QFI of a non-full-rank state
is completely determined by its support, i.e., the subset of $\{|\Phi
_{n}\rangle\langle\Phi_{n}|\}$ with nonzero eigenvalues. For a two-level
system, its density matrix can always be expressed in terms of the Pauli
matrices $\hat{\boldsymbol{\sigma}}$ as $\hat{\rho}_{\zeta}=(1/2)(1+\hat
{\boldsymbol{\sigma}}\cdot\mathbf{n})$, where $\mathbf{n}\equiv
\operatorname*{Tr}\hat{\boldsymbol{\sigma}}\hat{\rho}_{\zeta}$ is the Bloch
vector. The QFI for such a state is
\cite{DittmannJPA1999,ZhongPRA2013,LiPRA2015}
\begin{equation}
\mathcal{F}=|\partial_{\zeta}\mathbf{n}|^{2}+\frac{(\partial_{\zeta
}|\mathbf{n}|^{2})^{2}}{4(1-|\mathbf{n}|^{2})},\label{QFI_TLS}%
\end{equation}
where the second term is absent when $|\mathbf{n}|=1$, i.e., when $\hat{\rho
}_{\zeta}$ is a pure state. When $\hat{\rho}_{\zeta}=\hat{\rho}_{\zeta}%
^{(1)}\otimes\cdots\otimes\hat{\rho}_{\zeta}^{(N)}$ is the direct product
state of $N$ quantum systems, its QFI is additive:\ $\mathcal{F}=\sum
_{n=1}^{N}\mathcal{F}^{(n)}$, where $\mathcal{F}^{(n)}$ is the QFI\ of
$\hat{\rho}_{\zeta}^{(n)}$.

Physically, the QFI measures the rate of variation of $\hat{\rho}_{\zeta}$
with the parameter $\zeta$, e.g., if we regard $\hat{\rho}_{\zeta}$ and
$\hat{L}_{\zeta}$ as classical variables, then $L_{\zeta}=\partial_{\zeta}%
\ln\rho_{\zeta}$ and Eq. (\ref{QFI_DEF}) becomes the average of $(\partial
_{\zeta}\ln\rho_{\zeta})^{2}$ over the state $\rho_{\zeta}$. Moreover, the
Bures distance between two quantum states $\hat{\rho}_{1}$ and $\hat{\rho}%
_{2}$ is defined as \cite{BuresTAMS1969}%
\[
\mathcal{D}(\hat{\rho}_{1},\hat{\rho}_{2})\equiv\sqrt{2}(1-\operatorname*{Tr}%
\sqrt{\sqrt{\hat{\rho}_{1}}\hat{\rho}_{2}\sqrt{\hat{\rho}_{1}}})^{1/2},
\]
where the second term $\operatorname*{Tr}\sqrt{\cdots}$ on the right-hand side
is the so-called Uhlmann fidelity \cite{UhlmannRMP1976}. For neighboring
states $\hat{\rho}_{\zeta}$ and $\hat{\rho}_{\zeta+d\zeta}$, the Bures
distance reduces to
\[
\mathcal{D}(\hat{\rho}_{\zeta},\hat{\rho}_{\zeta+d\zeta})=\frac{1}{2}%
\sqrt{\mathcal{F}}d\zeta,
\]
so the QFI measures the distinguishability between two neighboring states
parametrized by $\zeta$.

The importance of the QFI for parameter estimation is manifested in the
inequalities Eqs. (\ref{FCFQ}) and (\ref{CRB}). Namely, given $\hat{\rho
}_{\zeta}$ and hence $\mathcal{F}(\zeta)$, the precision of \textit{any}
unbiased estimator from $N$ repetitions of \textit{any} measurement is limited
by the inequality%
\begin{equation}
\delta\zeta\geq\frac{1}{\sqrt{N\mathcal{F}(\zeta)}},\label{QCRB}%
\end{equation}
known as the quantum Cram\'{e}r-Rao bound
\cite{HelstromBook1976,BraunsteinPRL1994}. Saturating this bound requires
saturating Eqs. (\ref{FCFQ}) and (\ref{CRB}) simultaneously, i.e., using
optimal measurements to convert all the QFI into the CFI and using optimal
unbiased estimators to convert all the CFI into the precision of the estimator.

\section{Measurement:\ classical Fisher information}

\label{SEC_MEASUREMENT}

A general measurement with discrete outcomes $\{u\}$ is described by the
positive-operator valued measure (POVM) elements $\{\hat{M}_{u}\}$ satisfying
the completeness relation $\sum_{u}\hat{M}_{u}^{\dagger}\hat{M}_{u}=1$. Given
a quantum state $\hat{\rho}_{\zeta}$, it yields an outcome $u$ according to
the probability distribution $P(u|\zeta)\equiv\operatorname*{Tr}\hat{M}%
_{u}\hat{\rho}_{\zeta}\hat{M}_{u}^{\dagger}$ that depends on $\zeta$. The
amount of information about $\zeta$ contained in each outcome is quantified by
the CFI \cite{KayBook1993}:%
\begin{equation}
F(\zeta)\equiv\sum_{u}P(u|\zeta)\left(  \frac{\partial\ln P(u|\zeta)}%
{\partial\zeta}\right)  ^{2}. \label{CFI_DEF}%
\end{equation}
For continuous outcomes, we need only replace $\sum_{u}$ by $\int du$
everywhere. Physically, the CFI quantifies the dependence of the measurement
distribution $P(u|\zeta)$ on the parameter $\zeta$. Actually, the Wootters'
distance \cite{WoottersPRD1981} between two probability distributions
$P^{(1)}(u)$ and $P^{(2)}(u)$ is%
\[
D(P^{(1)},P^{(2)})\equiv\cos^{-1}\left(  \sum_{u}\sqrt{P^{(1)}(u)P^{(2)}%
(u)}\right)  .
\]
For neighboring distributions $P^{(1)}(u)=P(u|\zeta)$ and $P^{(2)}%
(u)=P(u|\zeta+d\zeta)$, the Wootters' distance reduces to%
\[
D(P(u|\zeta),P(u|\zeta+d\zeta))=\frac{1}{2}\sqrt{F(\zeta)}d\zeta,
\]
so the CFI measures the distinguishability between neighboring measurement
distributions parametrized by $\zeta$.

Since the probability distribution function is the classical counterpart of
the quantum mechanical density matrix, the CFI (Wootters' distance) is the
classical counterpart of the QFI (Bures distance). The inequality Eq.
(\ref{FCFQ}) expresses the simple fact that no new information about $\zeta$
can be generated in the measurement process:\ optimal (non-optimal)
measurements convert all (part) of the QFI\ into the CFI. Given $\hat{\rho
}_{\zeta}$, the optimal measurement is \textit{not} unique. The projective
measurement on the symmetric logarithmic derivative operator $\hat{L}%
_{\zeta_{\mathrm{true}}}$ has been identified \cite{BraunsteinPRL1994} as an
optimal measurement, but $\zeta_{\mathrm{true}}$ is not known. To circumvent
this problem, the simplest way is to find other optimal measurements that do
not depend on $\zeta_{\mathrm{true}}$. Another solution
\cite{NielsenJPAMG2000} is to approximate $\hat{L}_{\zeta_{\mathrm{true}}}$ by
$\hat{L}_{\zeta_{\mathrm{est}}}$, where $\zeta_{\mathrm{est}}$ is our best
guess to $\zeta_{\mathrm{true}}$, i.e., the optimal unbiased estimator, as we
discuss below.

\section{Data processing:\ optimal unbiased estimators}

\label{SEC_MLE}

Given the measurement distribution $P(u|\zeta)$ and hence the CFI $F(\zeta)$
of each outcome, the precision $\delta\zeta$ of \textit{any} unbiased
estimator $\zeta_{\mathrm{est}}(\mathbf{u})$ constructed from the outcomes
$\mathbf{u}\equiv(u_{1},\cdots,u_{N})$ of $N$ repeated measurements is limited
by the Cram\'{e}r-Rao bound Eq. (\ref{CRB}), which expresses the simple fact
that no new information about $\zeta$ can be generated in the data
processing:\ optimal (non-optimal) unbiased estimators convert all (part) of
the CFI into the useful information $(\delta\zeta)^{-2}$ quantified by the
precision $\delta\zeta$. Finding optimal unbiased estimators is an important
step in parameter estimation. In the limit of large $N$, two kinds of
estimators are known to be unbiased and optimal: the maximum likelihood
estimator and the Bayesian estimator \cite{KayBook1993}, as we introduce now.

Before any measurements, our prior knowledge about the unknown parameter
$\zeta$ is quantified by certain probability distribution $P_{0}(\zeta)$,
e.g., a $\delta$-like distribution corresponds to knowing $\zeta$ exactly, a
flat distribution corresponds to completely no knowledge about $\zeta$, while
a Gaussian distribution $P_{0}(\zeta)\propto e^{-(\zeta-\zeta_{0}%
)^{2}/(2\sigma_{0}^{2})}$ corresponds to knowing $\zeta$ to be $\zeta_{0}$
with a typical uncertainty $\sigma_{0}$.

Upon getting the first outcome $u_{1}$, our knowledge about $\zeta$ is
immediately refined from $P_{0}(\zeta)$ to%
\[
P_{u_{1}}(\zeta)=\frac{P_{0}(\zeta)P(u_{1}|\zeta)}{\mathcal{N}(u_{1})}%
\]
according to the Bayesian rule \cite{ToussaintRMP2011}, where $\mathcal{N}%
(u_{1})\equiv\int d\zeta P_{0}(\zeta)P(u_{1}|\zeta)$ is a normalization factor
ensuring $P_{u_{1}}(\zeta)$ is normalized to unity: $\int P_{u_{1}}%
(\zeta)d\zeta=1$. Here $P_{u_{1}}(\zeta)$ is the \textit{posterior}
probability distribution of $\zeta$ conditioned on the outcome of the
measurement being $u_{1}$:\ its parametric dependence on $u_{1}$ means that
different measurement outcomes leads to different refinement of knowledge
about $\zeta$.

Upon getting the second outcome $u_{2}$, our knowledge is immediately refined
from $P_{u_{1}}(\zeta)$ to
\[
P_{u_{1}u_{2}}(\zeta)=\frac{P_{0}(\zeta)P(u_{1}|\zeta)P(u_{2}|\zeta
)}{\mathcal{N}(u_{1},u_{2})},
\]
where $\mathcal{N}(u_{1},u_{2})=\int P_{0}(\zeta)P(u_{1}|\zeta)P(u_{2}%
|\zeta)d\zeta$ is a normalization factor for the posterior distribution
$P_{u_{1}u_{2}}(\zeta)$. If we omit the trivial normalization factors, then
the measurement-induced knowledge refinement becomes
\[
P_{0}(\zeta)\overset{u_{1}}{\longrightarrow}P_{0}(\zeta)P(u_{1}|\zeta
)\overset{u_{2}}{\longrightarrow}P_{0}(\zeta)P(u_{1}|\zeta)P(u_{2}%
|\zeta)\overset{u_{3}}{\longrightarrow}\cdots.
\]

Upon getting $N$ outcomes $\mathbf{u}\equiv(u_{1},\cdots,u_{N})$, our
knowledge about $\zeta$ is quantified by the posterior distribution
\[
P_{\mathbf{u}}(\zeta)\sim P_{0}(\zeta)P(\mathbf{u}|\zeta)
\]
up to a trivial normalization factor, where $P(\mathbf{u}|\zeta)=P(u_{1}%
|\zeta)\cdots P(u_{N}|\zeta)$ is the probability for getting the outcome
$\mathbf{u}$. The posterior distribution $P_{\mathbf{u}}(\zeta)$ completely
describe our state of knowledge about $\zeta$. Nevertheless, sometimes a
single number, i.e., an unbiased estimator, is required as the best guess to
$\zeta_{\mathrm{true}}$. There are two well-known estimators: the maximum
likelihood estimator \cite{KayBook1993}
\begin{equation}
\zeta_{\mathrm{M}}(\mathbf{u})\equiv\arg\max P_{\mathbf{u}}(\zeta) \label{MLE}%
\end{equation}
is the peak position of $P_{\mathbf{u}}(\zeta)$ as a function of $\zeta$,
while the Bayesian estimator \cite{KayBook1993}
\begin{equation}
\zeta_{\mathrm{B}}(\mathbf{u})\equiv\int\zeta P_{\mathbf{u}}(\zeta
)d\zeta\label{BAYESIAN}%
\end{equation}
is the average of $\zeta$. For large $N$, both estimators are unbiased and
optimal: $\langle\zeta_{\alpha}\rangle=\zeta$ and $\delta\zeta_{\alpha
}=1/\sqrt{NF(\zeta)}$, where $\alpha=\mathrm{M}$ or $\mathrm{B}$, and
$\langle\cdots\rangle$ denotes the average over a large number of estimators
obtained by repeating the $N$-outcome estimation scheme many times and
$\delta\zeta_{\alpha}$ is defined as Eq. (\ref{UNCERTAINTY}) or%
\begin{equation}
\delta\zeta_{\alpha}=\sqrt{\int[\zeta-\zeta_{\alpha}(\mathbf{u})]^{2}%
P_{\mathbf{u}}(\zeta)d\zeta}. \label{DZ_DEF}%
\end{equation}
For a simple understanding, we consider $N\rightarrow\infty$, so the number of
occurrence of a specific outcome $u$ approaches $NP(u|\zeta_{\mathrm{true}})$.
Then, up to a trivial normalization factor, the posterior distribution
$P_{\mathbf{u}}(\zeta)$ approaches
\[
\prod_{u}[P(u|\zeta)]^{NP(u|\zeta_{\mathrm{true}})}=\exp\left(  N\sum
_{u}P(u|\zeta_{\mathrm{true}})\ln P(u|\zeta)\right)  ,
\]
which exhibits a sharp peak at $\zeta=\zeta_{\mathrm{true}}$. For large $N$,
$P_{\mathbf{u}}(\zeta)$ is nonzero only in the vicinity of $\zeta
_{\mathrm{true}}$. This justifies a Taylor expansion around $\zeta
_{\mathrm{true}}$, leading to the Gaussian form $P_{\mathbf{u}}(\zeta)\sim
e^{-(\zeta-\zeta_{\mathrm{true}})^{2}/(2\sigma^{2})}$ with a standard
deviation $\sigma\equiv1/\sqrt{NF(\zeta_{\mathrm{true}})}$. Then we have
$\zeta_{\mathrm{M}}=\zeta_{\mathrm{B}}=\zeta_{\mathrm{true}}$ and $\delta
\zeta_{\mathrm{M}}=\delta\zeta_{\mathrm{B}}=\sigma$, so both estimators are
unbiased and optimal in the large $N$ limit.

Usually, calculating $\zeta_{\mathrm{M}}$ [Eq. (\ref{MLE})] or $\zeta
_{\mathrm{B}}$ [Eq.\ (\ref{BAYESIAN})] requires intensive computational costs.
The situation simplifies when the data comes from binary-outcome measurements,
i.e., measurements that yield only two possible outcomes (denoted by $+$ and
$-$)\ according to the probability distribution $P(\pm|\zeta)$. In this case,
let $N_{+}$ ($N_{-}$) denote the number of outcome $+$ (outcome $-$) from the
$N$ measurements, then solving $P(+|\zeta)-P(-|\zeta)=(N_{+}-N_{-})/N$ for
$\zeta$ gives a simple estimator that is unbiased and optimal for large $N$
\cite{FengPRA2014}. In this work, we always adopt the maximum likelihood estimator.

\section{An example}

\label{SEC_EXAMPLE}

\begin{figure}[ptb]
\includegraphics[width=\columnwidth]{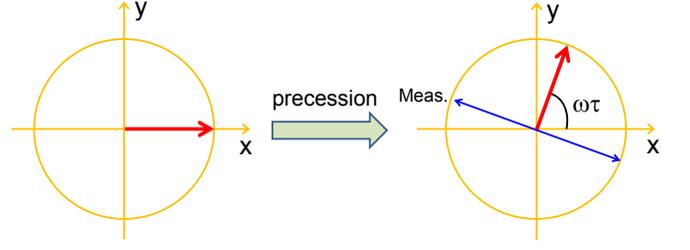}\caption{(Color online). Free
precession of a spin-1/2 around the $z$ axis by an angle $\omega\tau$. The red
arrows denote the initial and final spin orientation and the blue arrow
denotes the optimal measurement axis determined by the symmetric logarithmic
derivative operator $\hat{L}_{\omega_{\mathrm{true}}}$.}%
\label{G_EXAMPLE}%
\end{figure}

Here we follow the three standard steps outlined in Fig. \ref{G_FRAMEWORK} to
estimate the level splitting $\omega$ of a spin-1/2 Hamiltonian
\[
\hat{H}_{0}=\frac{1}{2}\omega\hat{\sigma}_{z}%
\]
by monitoring its free precession. For convenience we define $\mathbf{e}%
_{\varphi}$ as a unit vector in the $xy$ plane with azimuth $\varphi$.

For step 1, we assume the initial state to be $|\psi_{\mathrm{in}}\rangle
=\cos(\Theta/2)|\uparrow\rangle+e^{i\Phi}\sin(\Theta/2)|\downarrow\rangle$,
where the controlling parameters $\Theta$ and $\Phi$ are to be optimized.
Next, the spin-1/2 undergoes $\omega$-dependent free precession for an
interval $\tau$ into the final state $|\psi_{\omega}\rangle=e^{-i\omega
\tau\hat{\sigma}_{z}/2}|\psi_{\mathrm{in}}\rangle$. The QFI\ in the final
state is calculated by Eq. (\ref{QFI_PURE}) as
\[
\mathcal{F}=\tau^{2}\sin^{2}\Theta,
\]
which is independent of $\Phi$. To maximize $\mathcal{F}$, we set $\Theta
=\pi/2$, and leave $\Phi$ arbitrary, i.e., any initial state whose average
spin lies in the $xy$ plane is optimal. For specificity, we set $\Phi=0$, so
the initial state is the $\hat{\sigma}_{x}=+1$ eigenstate and the final state
is the $\hat{\boldsymbol{\sigma}}\cdot\mathbf{e}_{\omega\tau}=+1$ eigenstate,
as shown in Fig. \ref{G_EXAMPLE}. Interestingly, the QFI can be increased
indefinitely by increasing the evolution time $\tau$, indicating the time as a
valuable quantum resource.

For step 2, we need to find optimal measurements to convert all the QFI into
the CFI. There are two ways to find optimal measurements. The first one is to
use the general conclusion \cite{BraunsteinPRL1994} that the projective
measurement on the symmetric logarithmic derivative operator $\hat{L}%
_{\omega_{\mathrm{true}}}$ is optimal (see Appendix B). Since the final state
is pure, we have
\[
\hat{L}_{\omega}=2\partial_{\omega}(|\psi_{\omega}\rangle\langle\psi_{\omega
}|)=\tau\boldsymbol{\hat{\sigma}}\cdot\mathbf{e}_{\pi/2+\omega\tau},
\]
i.e., measuring $\boldsymbol{\hat{\sigma}}\cdot\mathbf{e}_{\pi/2+\omega
_{\mathrm{true}}\tau}$ (blue arrow in Fig. \ref{G_EXAMPLE}) is optimal.
Actually, this measurement gives two possible outcomes $\pm1$ according to the
distribution $P(\pm1|\omega)=[1\pm\sin(\omega-\omega_{\mathrm{true}})\tau]/2$
and the CFI\ is computed from Eq. (\ref{CFI_DEF}) as $F(\omega)|_{\omega
\rightarrow\omega_{\mathrm{true}}}=\tau^{2}=\mathcal{F}$. Physically, this
amounts to measuring the spin-1/2 along an axis (blue arrow in Fig.
\ref{G_EXAMPLE}) perpendicular to the spin orientation of the final state
$|\psi_{\omega}\rangle$ (red arrow in Fig. \ref{G_EXAMPLE}) to maximize the
dependence of the measurement distribution on the parameter $\omega$. However,
since $\omega_{\mathrm{true}}$ is unknown, complicated adaptive measurements
are necessary. The second method is to consider a projective measurement on
the spin-1/2 along a general axis parametrized by polar angle $\theta$ and
azimuth $\varphi$, which gives the measurement distribution $P(\pm
1|\omega)=[1\pm\sin\theta\cos(\omega\tau-\varphi)]/2$ and hence the CFI
\[
F=\tau^{2}\frac{\sin^{2}\theta\sin^{2}(\omega\tau-\varphi)}{1-\sin^{2}%
\theta\cos^{2}(\omega\tau-\varphi)}.
\]
To maximize $F$, we set $\theta=\pi/2$, then $F=\tau^{2}$ attains the QFI,
i.e., measuring the spin-1/2 along an arbitrary axis in the $xy$ plane form a
family of optimal measurements. For specificity we set $\varphi=0$,
corresponding to measuring $\hat{\sigma}_{x}$.

For step 3, suppose we have no prior knowledge about $\omega$ before the
measurements. Next we repeat the initialization-evolution-measurement cycle
twice and obtain two outcomes $\mathbf{u}\equiv(u_{1},u_{2})$. Upon getting
these outcomes, our knowledge about $\omega$ is immediately refined to (up to
a constant normalization factor) $P_{\mathbf{u}}(\omega)\sim P(u_{1}%
|\omega)P(u_{2}|\omega)$, e.g., if both outcomes are $+1$, then $P_{\mathbf{u}%
}(\omega)\sim\cos^{4}(\omega\tau/2)$ shows many peaks at integer multiples of
$2\pi/\tau$, corresponding to an infinite number of maximum likelihood
estimators $\omega_{\mathrm{M}}=2n\pi/\tau$ ($n\in\mathbb{Z})$. If both
outcomes are $-1$, then $P_{\mathbf{u}}(\omega)\sim\sin^{4}(\omega\tau/2)$
shows maxima at odd multiples of $\pi/\tau$, corresponding to $\omega
_{\mathrm{M}}=(2n+1)\pi/\tau$ ($n\in\mathbb{Z})$. If one outcome is $+1$ and
the other is $-1$, then $P_{\mathbf{u}}(\omega)\sim\sin(\omega\tau)$ and
$\omega_{\mathrm{M}}=(2n+1)\pi/(2\tau)$ ($n\in\mathbb{Z})$. In any case, the
maximum likelihood estimator is not unique, because the measurement
distribution $P(u|\omega)$ is an even function of $\omega$ with a period
$2\pi/\tau$, so $\omega$ and $-\omega$ (or $\omega$ and $\omega+2\pi/\tau$)
give exactly the same measurement distribution and hence cannot be
distinguished. Such ambiguity can be eliminated by combining the information
gained from measurements with different evolution time $\tau$
\cite{SaidPRB2011,SergeevichPRA2011}.

\section{Feedback: adaptive measurement protocols}

\label{SEC_FEEDBACK}

As mentioned before, usually the CFI depends on the parameter $\zeta$ to be
estimated, so optimizing the initial state, the evolution process, and the
measurement scheme requires knowledge about $\zeta$, which is unknown. A
standard solution is adaptive measurement protocols: after each
initialization-evolution-measurement cycle, the measurement outcome is
immediately used to refine our knowledge about $\zeta$, which in turn is used
to optimize the next cycle, as shown in Fig. \ref{G_FRAMEWORK}.

There are two categories of adaptive protocols. The first category focuses on
maximizing the CFI \cite{OlivaresJPB2009,BrivioPRA2010,PangNC2017}. Suppose
the CFI\ $F(\zeta,\boldsymbol{\theta})$ depends on $\zeta$ and some parameters
$\boldsymbol{\theta}$ that control the initialization, evolution, and
measurement processes. The simplest idea is to tune $\boldsymbol{\theta}$ to
maximize the $F(\zeta,\boldsymbol{\theta})$. However, usually the optimal
$\boldsymbol{\theta}$ leading to maximal CFI\ depends on the unknown parameter
$\zeta$. Suppose at the end of the $(n-1)$th
initialization-evolution-measurement cycle, our knowledge is quantified by a
distribution $P(\zeta)$, then a natural solution is to choose
$\boldsymbol{\theta}$ in the $n$th cycle to maximize the CFI averaged over the
distribution of $\zeta$, i.e.,\
\begin{equation}
\bar{F}(\boldsymbol{\theta})\equiv\int F(\zeta,\boldsymbol{\theta}%
)P(\zeta)d\zeta\approx F(\zeta_{\mathrm{M}},\boldsymbol{\theta}), \label{FAVE}%
\end{equation}
where the second step is valid when the maximum of $P(\zeta)$ at
\begin{equation}
\zeta_{\mathrm{M}}\equiv\arg\max P(\zeta) \label{MLE_PREVIOUS}%
\end{equation}
is very sharp compared with $F(\zeta,\boldsymbol{\theta})$.

The second category focuses on optimizing the expected information gain from
the estimator \cite{BerryPRL2000,BerryPRA2001,SaidPRB2011,SergeevichPRA2011}.
At the end of the $(n-1)$th cycle, our knowledge about $\zeta$ is quantified
by the distribution $P(\zeta)$. In the $n$th cycle with the controlling
parameters $\boldsymbol{\theta}$, the measurement distribution is
$P_{\boldsymbol{\theta}}(u|\zeta)$, which depend on $\zeta$ and
$\boldsymbol{\theta}$. If the measurement outcome of this cycle is $u$, then
our knowledge about $\zeta$ would be updated to the distribution
\begin{equation}
P_{u,\boldsymbol{\theta}}(\zeta)\sim P(\zeta)P_{\boldsymbol{\theta}}%
(u|\zeta),\label{PU_THETA}%
\end{equation}
the maximum likelihood estimator would be $\zeta_{\mathrm{M}}%
(u,\boldsymbol{\theta})\equiv\arg\max P_{u,\boldsymbol{\theta}}(\zeta)$, and
its uncertainty $\delta\zeta(u,\boldsymbol{\theta})$ would be given by Eq.
(\ref{DZ_DEF}) with $P_{\mathbf{u}}(\zeta)\rightarrow P_{u,\boldsymbol{\theta
}}(\zeta)$ and $\zeta_{\alpha}(\mathbf{u})\rightarrow\zeta_{\mathrm{M}%
}(u,\boldsymbol{\theta})$. Since the probability for this outcome $u$ to occur
is given by $P_{\boldsymbol{\theta}}(u|\zeta)$ averaged over the distribution
$P(\zeta)$, i.e., $P_{\boldsymbol{\theta}}(u)\equiv\int P_{\boldsymbol{\theta
}}(u|\zeta)P(\zeta)d\zeta$, we should choose $\boldsymbol{\theta}$ in the
$n$th cycle to minimize the expected uncertainty:\
\begin{equation}
\overline{\delta\zeta}(\boldsymbol{\theta})\equiv\sum_{u}P_{\boldsymbol{\theta
}}(u)\delta\zeta(u,\boldsymbol{\theta}).\label{DZ_AVE}%
\end{equation}
When the maximum of $P(\zeta)$ at $\zeta_{\mathrm{M}}$ [Eq.
(\ref{MLE_PREVIOUS})] is very sharp, we have $P_{\boldsymbol{\theta}%
}(u)\approx P_{\boldsymbol{\theta}}(u|\zeta_{\mathrm{M}})$, so \
\begin{equation}
\overline{\delta\zeta}(\boldsymbol{\theta})\approx\sum_{u}%
P_{\boldsymbol{\theta}}(u|\zeta_{\mathrm{M}})\delta\zeta(u,\boldsymbol{\theta
}).\label{BERRY}%
\end{equation}
The key idea of this adaptive scheme is to optimize the controlling parameters
of the next cycle to minimize the expected uncertainty at the end of that
cycle [Eq. (\ref{DZ_AVE}) or (\ref{BERRY})], so it is known as the locally
optimal adaptive scheme \cite{BerryPRL2000,BerryPRA2001}. A straightforward
extension is to optimize simultaneously the $M$ controlling parameters of the
next $M$ cycles to minimize the expected uncertainty at the end of these
cycles. Increasing $M$ improves the performance at the cost of exponentially
increasing computational cost \cite{BerryPRA2009}, so the $M=1$ scheme is the
most widely used one in Hamiltonian parameter estimation
\cite{BerryPRL2000,BerryPRA2001,SaidPRB2011,SergeevichPRA2011}.

Although both the CFI-based adaptive scheme and the locally optimal adaptive
scheme have been widely used, their connection remains unclear. Here we prove
their equivalence in the limit of very accurate knowledge $P(\zeta)$ at the
beginning of the $n$th initialization-evolution-measurement cycle, as
quantified by a sharp Gaussian distribution $P(\zeta)=e^{-(\zeta
-\zeta_{\mathrm{M}})^{2}/(2\sigma^{2})}/(\sqrt{2\pi}\sigma)$ with a small
standard deviation $\sigma$. This justifies a Taylor expansion of Eq.
(\ref{PU_THETA}) around $\zeta_{\mathrm{M}}$, which gives
$P_{u,\boldsymbol{\theta}}(\zeta)\sim e^{-(\zeta-\zeta_{u,\boldsymbol{\theta}%
})^{2}/(2\sigma_{u,\boldsymbol{\theta}}^{2})}$ with%
\[
\sigma_{u,\boldsymbol{\theta}}^{-2}=\sigma^{-2}-\left(  \frac{\partial^{2}\ln
P_{\boldsymbol{\theta}}(u|\zeta)}{\partial\zeta^{2}}\right)  _{\zeta
_{\mathrm{M}}}\Rightarrow\sigma_{u,\boldsymbol{\theta}}\approx\sigma
+\frac{\sigma^{3}}{2}\left(  \frac{\partial^{2}\ln P_{\boldsymbol{\theta}%
}(u|\zeta)}{\partial\zeta^{2}}\right)  _{\zeta_{\mathrm{M}}}.
\]
Then we have $\zeta_{\mathrm{M}}(u,\boldsymbol{\theta})=\zeta
_{u,\boldsymbol{\theta}}$ and $\delta\zeta(u,\boldsymbol{\theta}%
)=\sigma_{u,\boldsymbol{\theta}}$, so the expected uncertainty in Eq.
(\ref{BERRY}) becomes%
\[
\overline{\delta\zeta}(\boldsymbol{\theta})\approx\sigma-\frac{1}{2}\sigma
^{3}F(\zeta_{\mathrm{M}},\boldsymbol{\theta}),
\]
thus minimizing $\overline{\delta\zeta}(\boldsymbol{\theta})$ amounts to
maximizing $F(\zeta_{\mathrm{M}},\boldsymbol{\theta})$ in Eq. (\ref{FAVE}).


\end{document}